\documentclass[aps,prd,
floatfix,preprintnumbers,superscriptaddress,nofootinbib,nameinlink,capitalise]{revtex4-2}

\usepackage[left=2.54cm, right=2.54cm, top=2.54cm, bottom=2.54cm]{geometry}
\usepackage{palatino}
\usepackage{amsmath, amssymb}
\usepackage{graphicx}
\usepackage{array, booktabs, multirow}
\usepackage{float}
\usepackage{hyperref}
\usepackage{subfig}
\usepackage{nameref}      
\usepackage{cleveref}
\usepackage{placeins}
\usepackage{verbatim}
\usepackage{textgreek}
\usepackage[justification=raggedright,singlelinecheck=false]{caption}
\usepackage{color}
\newcommand{\bmat}{\left(\begin{array}}
\newcommand{\emat}{\end{array}\right)}
\newcommand{\be}{\begin{equation}}
\newcommand{\ee}{\end{equation}}
\newcommand{\bea}{\begin{eqnarray}}
\newcommand{\eea}{\end{eqnarray}}

\usepackage{graphicx}
\usepackage{subcaption}

\begin{document}

\title{Radiatively Corrected Starobinsky Inflation and Primordial Gravitational Waves in Light of ACT Observations}

\author{Waqas Ahmed}
\email{waqasmit@hbpu.edu.cn}
\affiliation{Center for Fundamental Physics and School of Mathematics and Physics, Hubei Polytechnic University, Huangshi 435003, China}

\author{Mansoor Ur Rehman}
\email{mansoor@qau.edu.pk}
\affiliation{Department of Physics, Faculty of Science, Islamic University of Madinah, 42351 Madinah, Saudi Arabia}

\noaffiliation


\vspace{1em}
\begin{abstract}
Nonminimal coupling between the inflaton and the Ricci scalar plays a crucial role in shaping the predictions of single-field inflationary models. While a quartic potential with such coupling represents one of the simplest realizations compatible with cosmological observations, it generically receives important radiative corrections when the inflaton interacts with other fields, particularly those involved in the reheating process. In this work, we focus on radiative corrections arising from bosonic scalar couplings and study their impact on inflationary dynamics within the nonminimally coupled quartic potential framework. We demonstrate that bosonic corrections, unlike fermionic ones, yield predictions more compatible with the latest constraints from the Atacama Cosmology Telescope (ACT) Data Release 6, especially when combined with Planck and BICEP/Keck data.
We begin with a general model involving a gauge-singlet real scalar inflaton coupled to a complex scalar field, which could be the Standard Model Higgs or a GUT Higgs. As a concrete realization, we also investigate the case where the inflaton serves as a dark matter candidate through a Higgs portal interaction while highlighting its potential to generate observable levels of primordial gravitational waves. Notably, the nonminimally coupled inflation model studied here is field-theoretically equivalent to the Starobinsky model, and the inclusion of quantum corrections from scalar fields leads to characteristic imprints in the $(n_s, r)$ plane, allowing for refined constraints on scalar couplings from current and future observations.
\end{abstract}

\maketitle

\section{Introduction}

Cosmic inflation, a period of accelerated expansion in the early universe, has become the cornerstone of modern cosmology. It provides elegant solutions to the horizon, flatness, and monopole problems, and it offers a mechanism for generating the primordial density perturbations that seeded the large-scale structure observed today \cite{Guth:1980zm, Linde:1981mu,Albrecht:1982wi}. Among the plethora of inflationary models, those incorporating non-minimal couplings between the inflaton scalar field and gravity have garnered significant interest \cite{Salopek:1988qh,Spokoiny:1984bd,Futamase:1987ua,Fakir:1990eg,Kaiser:1994vs,Komatsu:1999mt,Nozari:2007eq,Park:2008hz}. Such couplings, typically of the form $\xi \phi^2 R$, arise naturally in numerous theoretical contexts. The dimensionless parameter $\xi$ governs the strength of the interaction between the scalar field $\phi$ and the Ricci scalar $R$, and plays a crucial role in flattening the potential in the Einstein frame, thereby enabling successful slow-roll inflation.
A prominent realization of this idea is the Standard Model Higgs inflation scenario \cite{Bezrukov:2007ep}, which has since been extended to a range of frameworks beyond the Standard Model. These include both non-supersymmetric models \cite{Okada:2010jf,Okada:2011en,Borah:2020wyc} and supersymmetric frameworks 
\cite{Einhorn:2009bh,Ferrara:2010yw,Lee:2010hj,Ferrara:2010in,Linde:2011nh,Ellis:2013xoa,Pallis:2014cda}. In particular, implementations using Higgs fields from supersymmetric grand unified theories (GUTs) have been extensively studied in the context of supergravity and no-scale models  \cite{Arai:2011nq,Pallis:2011gr,Leontaris:2016jty,Okada:2017rbf,Ellis:2017jcp,Ahmed:2018jlv,Masoud:2019cen,Abid:2021jvn,Ahmed:2021dvo,Okada:2022yvq,Ahmed:2023rky,Ijaz:2023cvc,Ijaz:2024zma}.

The Starobinsky inflation model \cite{Starobinsky:1980te}, originally formulated through higher-curvature gravity, can be recast into a scalar-tensor theory exhibiting such a non-minimal coupling. Its simplicity and robust predictions have made it a leading candidate consistent with precision cosmological observations. However, the tree-level potential often employed in such models neglects the important effects of quantum corrections induced by interactions between the inflaton and other fields. In realistic particle physics setups, the inflaton couples to additional scalar (bosonic) and fermionic degrees of freedom, generating loop corrections \cite{Coleman:1973jx} that can significantly alter the inflationary dynamics and the resulting cosmological observables \cite{Senoguz:2008nok,Rehman:2009wv,Okada:2010jf,Rehman:2010es,Ahmed:2014cma}.
These radiative corrections introduce competing effects: bosonic loops tend to increase the inflationary observables such as the scalar spectral index \(n_s\) and the tensor-to-scalar ratio \(r\), while fermionic loops typically contribute with the opposite sign \cite{Okada:2010jf,Bostan:2019fvk}. Understanding this interplay is crucial for making accurate theoretical predictions and for interpreting observational data within well-motivated particle physics frameworks.

Recent advances in observational cosmology provide an unprecedented opportunity to test and constrain these refined inflationary models. Notably, the Atacama Cosmology Telescope (ACT) Data Release 6 (DR6) \cite{ACT:2025tim}, combined with data from the Planck 2018 mission and BICEP/Keck 2018 observations \cite{Planck:2018jri,Planck:2018vyg}, has yielded the most precise measurements to date of the cosmic microwave background (CMB) temperature and polarization anisotropies. These datasets have slightly revised the estimated values of key inflationary parameters, including the scalar spectral index \(n_s\), prompting renewed scrutiny of inflationary potentials beyond the minimal scenarios.

In this context, the present study undertakes a detailed analysis of the radiatively corrected Starobinsky (or nonminimal) inflation model, aiming to bridge theoretical predictions with the latest observational constraints. We study a quartic inflaton potential non-minimally coupled to gravity, incorporating radiative corrections arising from bosonic and fermionic loops. Using a conformal transformation, the action is recast into the Einstein frame with canonical gravity, facilitating the study of inflationary dynamics and the computation of observables.
We begin by demonstrating that the minimal $\phi^4$ inflationary model is disfavored by current cosmological observations, in the absence of quantum corrections. Introducing a small but non-zero non-minimal coupling to gravity, specifically $\xi \gtrsim 0.02$, enables the model to generate inflationary predictions compatible with the Planck and ACT datasets for a standard choice of 60 $e$-folds. However, this compatibility is confined to a narrow region in parameter space, reflecting the restrictive nature of observational constraints on the model in its classical form.

When radiative corrections are incorporated, the inflationary dynamics are significantly modified. In particular, the bosonic loop contributions tend to raise both the scalar spectral index $n_s$ and the tensor-to-scalar ratio $r$, effectively extending the viable parameter space. These corrections shift the theoretical predictions enters the region favored by Planck and ACT datasets. The resulting predictions not only satisfy the $2\sigma$ bounds but also fall comfortably within the $1\sigma$ confidence region of current data.

The structure of the paper is as follows: In Section~\ref{sec2}, we introduce the theoretical framework, including a generic model setup and the interaction Lagrangian involving scalar and fermionic fields. Section~~\ref{sec3} is devoted to the radiatively corrected potential in Jordan Frame. In Section ~\ref{sec4}, we present the transformation from the Jordan frame to the Einstein frame, including the derivation of the effective potential and the canonical normalization of the scalar field. Section ~\ref{sec5} provides details of the slow-roll parameters, while Section ~\ref{sec6} offers a comprehensive account of the numerical analysis. In Section~\ref{sec7}, we present a realistic scenario in which the inflaton also serves as a scalar dark matter candidate, analyzed using a renormalization group (RG) improved framework.
Finally, Section~\ref{sec:conclusion} summarizes our conclusions.

\section{Theoretical Framework}\label{sec2}

We consider the Jordan frame as our starting point, where the tree-level action involves a real scalar inflaton $\phi$ non-minimally coupled to gravity, supplemented by additional bosonic and fermionic degrees of freedom \cite{Kaiser:1994vs,Bezrukov:2007ep, Rubio:2018ogq, Okada:2010jf,Rehman:2010es}.
\bea
S_J^{\text{tree}} &=& \int d^4x \, \sqrt{-g} \Bigg[ 
 -\left( \frac{1}{2} m_P^2 + \frac{1}{2} \xi \phi^2
+ \xi_{H} H^{\dagger} H \right) \mathcal{R}  \\
&+& \frac{1}{2} (\partial_\mu \phi)(\partial^\mu \phi) 
- \frac{1}{2} m^2 \phi^2 
- \frac{\lambda_{\phi}}{4!} \phi^4 \\
&+& (\partial_\mu H)(\partial^\mu H)^{\dagger} 
- (\lambda_{\phi H}\phi^2 - m_H^2) H H^{\dagger} - \lambda_H (H^{\dagger} H)^2   \\
&+& i \overline{N} \gamma^{\mu} \partial_{\mu} N 
- \frac{1}{2} (y_{\phi} \phi + m_N) \overline{N^c} N + \cdots 
\Bigg].
\eea
Here, $m_P$ denotes the reduced Planck mass, $\xi$ ($\xi_{H}$) is the non-minimal coupling constant between the scalar field $\phi$ ($H$) and the Ricci scalar $\mathcal{R}$, and $\lambda_{\phi} >0$ ($\lambda_{H}>0$) is its quartic self-coupling. The complex scalar field $H$ has a quartic coupling to the inflaton $\phi$ with strength $\lambda_{\phi H}$, a typical interaction that arises not only in inflaton-dark matter unification models \cite{Lerner:2009xg,Okada:2010jd} but also in many beyond SM theories \cite{Rehman:2008qs,Okada:2011en,Lazarides:2022ezc}, including those that preserve classical scale invariance \cite{Hong:2025tyi,Hong:2025cae}. 
The right-handed neutrinos $N$ are SM gauge singlets and interact with $\phi$ via the Yukawa coupling $y_\phi$.
A field-dependent Majorana mass $(y_\phi \phi + m_N) \overline{N^c} N$ not only supports a seesaw mechanism for neutrino mass generation but also allows for reheating via lepton-number-violating processes relevant for baryogenesis via leptogenesis.

The complex scalar field $H$ may be identified with the Standard Model (SM) Higgs doublet or a Higgs field associated with an extended gauge symmetry, such as in a grand unified theory (GUT). During inflation, the inflaton field $\phi$ acquires large values, which in turn generate a sizable effective mass for $H$ through the Higgs portal interaction $\lambda_{\phi H} \phi^2 H^\dagger H$. This large mass dynamically drives $H$ to zero, allowing the inflationary dynamics to be well-approximated by a single-field scenario dominated by $\phi$. Despite being stabilized during inflation, the fields $H$ and the right-handed neutrinos $N_i$ play a crucial role by inducing radiative corrections to the inflaton potential via the Coleman-Weinberg mechanism.

\section{Radiatively Corrected Effective Potential in Jordan Frame}\label{sec3}
Radiative corrections to the scalar potential become important in determining the inflationary dynamics, especially in the presence of additional scalar and fermionic degrees of freedom. The above model provides a basic setting to explore how quantum corrections from both bosonic and fermionic fields influence the inflaton potential and inflationary observables.  Such loop corrections arise through vacuum fluctuations of $H$ and $N$ in the inflationary background and can be incorporated using the Coleman–Weinberg one-loop effective potential given by,
\bea
V_J(\phi)  & = & \frac{1}{2} m^2 \phi^2 + \frac{\lambda_\phi}{4!} \phi^4 \nonumber \\
& + & \frac{1}{64\pi^2} \Bigg[ 
\left(m^2 + \frac{\lambda_\phi}{2} \phi^2\right)^2  \ln\left( \frac{m^2 + \frac{\lambda_\phi}{2} \phi^2}{\mu^2} \right) \nonumber \\
& + & 2 \, \mathcal{N}_H \left(\lambda_{\phi H} \phi^2 - m_H^2\right)^2 \ln\left( \frac{\lambda_{\phi H} \phi^2 - m_H^2}{\mu^2} \right)  \nonumber \\
& - & 2 \, \mathcal{N}_N \left(y_\phi \phi + m_N\right)^4 \ln\left( \frac{(y_\phi \phi + m_N)^2}{\mu^2} \right)
\Bigg],
\eea
where $\mathcal{N}_H$ is the dimension of the complex Higgs field $H$ and $\mathcal{N}_H=3$ is the number of right-handed neutrino fields $N$. For large field values during inflation, $\lambda_{\phi} \phi^2 \gg m^2$, $\lambda_{\phi H } \phi^2 \gg m_H^2$ and $y_{\phi} \phi \gg m_N$, the one-loop corrected scalar potential takes the following form,
\begin{equation}\label{fullV}
V_J(\phi) \simeq 
\left[ \frac{\lambda_{\phi}}{4!} \phi^4 + \kappa_b \phi^4 \ln\left( \frac{\phi}{\mu_b} \right) - \kappa_f \phi^4 \ln\left( \frac{\phi}{\mu_f} \right) \right],
\end{equation}
where the bosonic and fermionic loop corrections are captured by
\be
\kappa_b = \frac{\mathcal{N}_H }{16 \pi^2} \lambda_{\phi H}^2, \quad
\kappa_f = \frac{\mathcal{N}_N}{16 \pi^2} y_\phi^4, \quad \text{with } \mu_b = \frac{\mu}{\sqrt{\lambda_{\phi H}}}  \text{ and } \mu_f = \frac{\mu}{y_{\phi}}.
\ee
\section{Einstein Frame and Field Redefinitions}\label{sec4}
To study inflationary dynamics in a frame with canonical gravity, we perform a conformal transformation from the Jordan frame to the Einstein frame. The conformal rescaling of the metric is defined as \cite{Maeda:1988ab}
\begin{equation}
g_{\mu\nu} \rightarrow g_{E\mu\nu} = \Omega^2(\phi) \, g_{\mu\nu},
\end{equation}
where the conformal factor $\Omega^2(\phi)$ is given by
\begin{equation}
\Omega^2(\phi) = 1 + \frac{\xi \phi^2}{m_P^2}.
\end{equation}
This transformation eliminates the non-minimal coupling between the scalar field and gravity, bringing the gravitational action to its canonical Einstein–Hilbert form \cite{Kaiser:1994vs}:
\begin{equation}
S_E = \int d^4 x \sqrt{-g_E} \left[ -\frac{1}{2} m_P^2 \mathcal{R}_E + \mathcal{L}_E(\phi, H, N) \right].
\end{equation}
However, the scalar sector now acquires a non-canonical kinetic term due to the field-dependent conformal factor. To recover canonical kinetic terms for the inflaton, we redefine the field $\phi$ in terms of a canonically normalized field $\sigma$ \cite{DeSimone:2008ei}:
\begin{equation}
(\sigma')^2 \equiv \left( \frac{d\sigma}{d\phi} \right)^2 = \frac{1 + \xi \phi^2 / m_P^2 + 6 \xi^2 \phi^2 / m_P^2}{\left(1 + \xi \phi^2 / m_P^2\right)^2}.
\label{eq:fieldredef}
\end{equation}
In the large-field regime ($\phi \gg m_P / \sqrt{\xi}$), which is relevant for inflation, this simplifies to:
\begin{equation}
\sigma(\phi) \approx \sqrt{\frac{3}{2}} m_P \ln \left( \frac{\xi \phi^2}{m_P^2}\right), \text{ or } \phi(\sigma) \approx \frac{m_P}{\sqrt{\xi}} \exp\left( \sqrt{\frac{1}{6}} \frac{\sigma}{m_P} \right).
\label{eq:largefield}
\end{equation}
The scalar potential in the Einstein frame is obtained by rescaling the Jordan frame potential by the conformal factor, yielding,
\begin{equation}
V_E(\phi) = \frac{V_J(\phi)}{\Omega^4(\phi)}.
\end{equation}
Incorporating the dominant one-loop radiative corrections from the scalar field $H$ and the fermion $N$, the effective potential in the Einstein frame takes the form:
\begin{equation}\label{fullV}
V_E(\phi) = \frac{1}{\left(1 + \frac{\xi \phi^2}{m_P^2}\right)^2}
\left[ \frac{\lambda_{\phi}}{4!} \phi^4 + \kappa_b \phi^4 \ln\left( \frac{\phi}{\mu_b} \right) - \kappa_f \phi^4 \ln\left( \frac{\phi}{\mu_f} \right) \right].
\end{equation}
As we demonstrate in the following sections, the bosonic corrections are particularly significant. They help lift the spectral index $n_s$ and bring the inflationary predictions into better agreement with the latest observational results from the Atacama Cosmology Telescope (ACT) \cite{ACT:2025tim}.

\section{Slow-roll Approximation}\label{sec5}
Before analyzing the dynamics of the model, it is instructive to revisit the standard slow-roll approximation used in inflationary cosmology. This formalism provides a simplified description of inflation driven by a scalar field slowly evolving down its potential.
The main slow-roll parameters in the Einstein frame are defined as follows:
\begin{align}
\epsilon(\phi) &= \frac{1}{2} m_P^2 \left( \frac{V_E'}{V_E \sigma'} \right)^2, \label{eq:epsilon} \\
\eta(\phi) &= m_P^2 \left( \frac{V_E''}{V_E (\sigma')^2} - \frac{V_E' \sigma''}{V_E (\sigma')^3} \right), \label{eq:eta} \\
\zeta^2(\phi) &= m_P^4 \left( \frac{V_E'}{V_E \sigma'} \right) \left( \frac{V_E'''}{V_E (\sigma')^3}
- 3 \frac{V_E'' \sigma''}{V_E (\sigma')^4} + 3 \frac{V_E' (\sigma'')^2}{V_E (\sigma')^5}
- \frac{V_E' \sigma'''}{V_E (\sigma')^4} \right). \label{eq:zeta}
\end{align}
Here, a prime denotes differentiation with respect to the scalar field $\phi$, and $\sigma'$ arises due to non-canonical normalization of the field in the Einstein frame.
The slow-roll approximation holds when
\[
\epsilon \ll 1, \quad |\eta| \ll 1, \quad \text{and} \quad \zeta^2 \ll 1.
\]
Under these conditions, the inflationary observables take the approximate forms:
\begin{align}
n_s &\simeq 1 - 6 \epsilon + 2 \eta, 
r \simeq 16 \epsilon,  \label{eq:ns-r} \\
\alpha_s \equiv \frac{d n_s}{d \ln k} &\simeq 16 \epsilon \eta - 24 \epsilon^2 - 2 \zeta^2. \label{eq:running}
\end{align}
The number of e-folds from a field value $\phi_0$ (corresponding to the pivot scale $k_0$) to the end of inflation $\phi_e$ is given by:
\begin{equation}
N_0 = \frac{1}{m_P} \int_{\phi_e}^{\phi_0} \frac{\sigma'}{\sqrt{2\epsilon(\phi)}} d\phi, \label{efold}
\end{equation}
where $\phi_e$ is determined by the condition
\[
\max\left[\epsilon(\phi_e), |\eta(\phi_e)|, \zeta^2(\phi_e)\right] = 1.
\]
The amplitude of the curvature (scalar) perturbation,  denoted as $A_s$ by Planck, can be expressed in terms of the slow-roll parameter $\epsilon$ as:
\begin{equation}
A_s(k_0) = \left. \frac{V_E/m_P^4}{24 \, \pi^2 \, \epsilon} \right|_{\phi = \phi_0}, 
\label{delta}
\end{equation}
where $A_s(k_0) = (2.19 \pm 0.06) \times 10^{-9}$ is the scalar amplitude at the pivot scale $k_0 = 0.05\,\mathrm{Mpc}^{-1}$, as reported in the recent combined analysis of ACT DR6 and Planck 2018 data~\cite{ACT:2025tim}.
For improved accuracy, our numerical evaluations incorporate the first-order corrections to the slow-roll expansion as derived in~\cite{Stewart:1993bc}, particularly for the spectral index $n_s$, the tensor-to-scalar ratio $r$, its running, and the power spectrum amplitude $A_s$. 

\begin{figure}[t]
    \centering
    \includegraphics[width=0.48\linewidth,height=6.7cm]{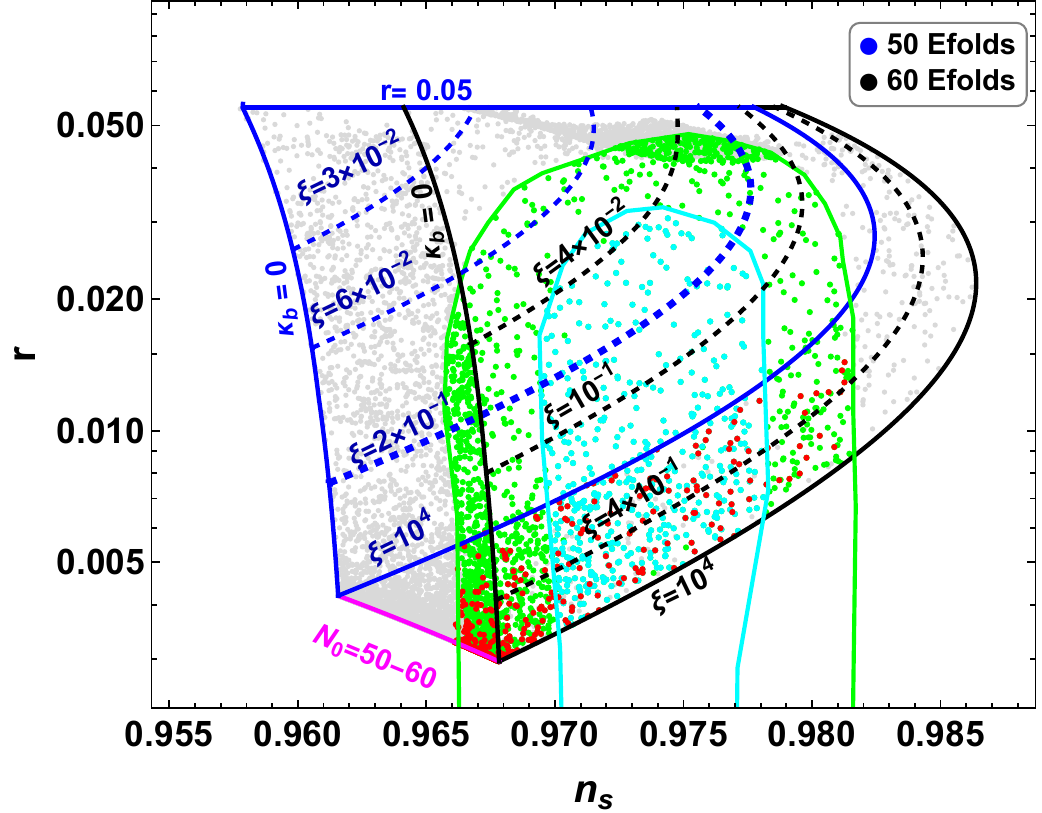}
    \quad
    \includegraphics[width=0.48\linewidth,height=6.7cm]{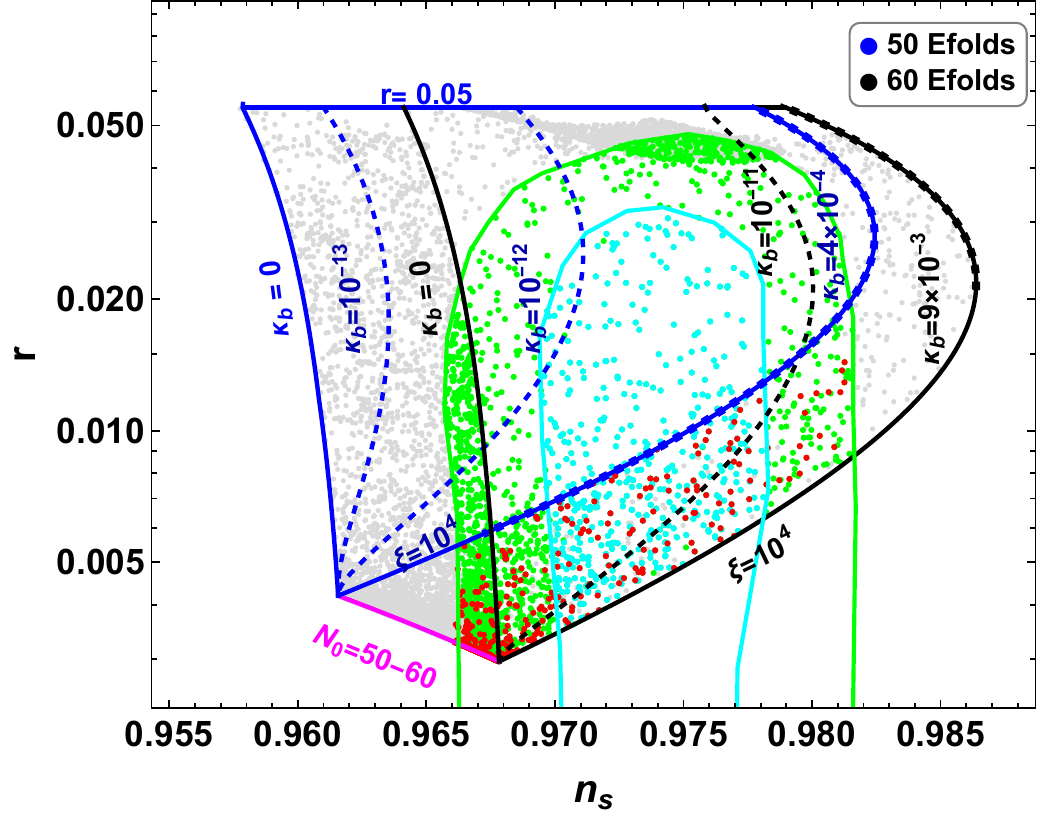}
      \quad
      \includegraphics[width=0.48\linewidth,height=6.7cm]{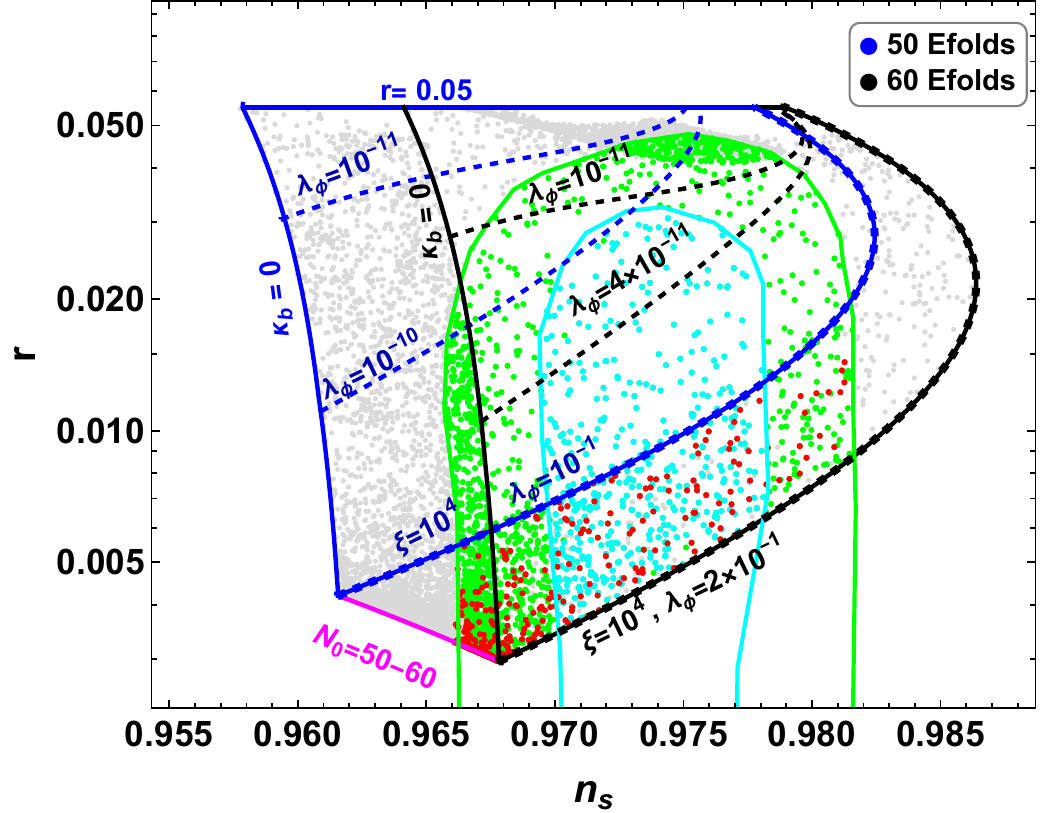}
      \quad
      \includegraphics[width=0.48\linewidth,height=6.7cm]{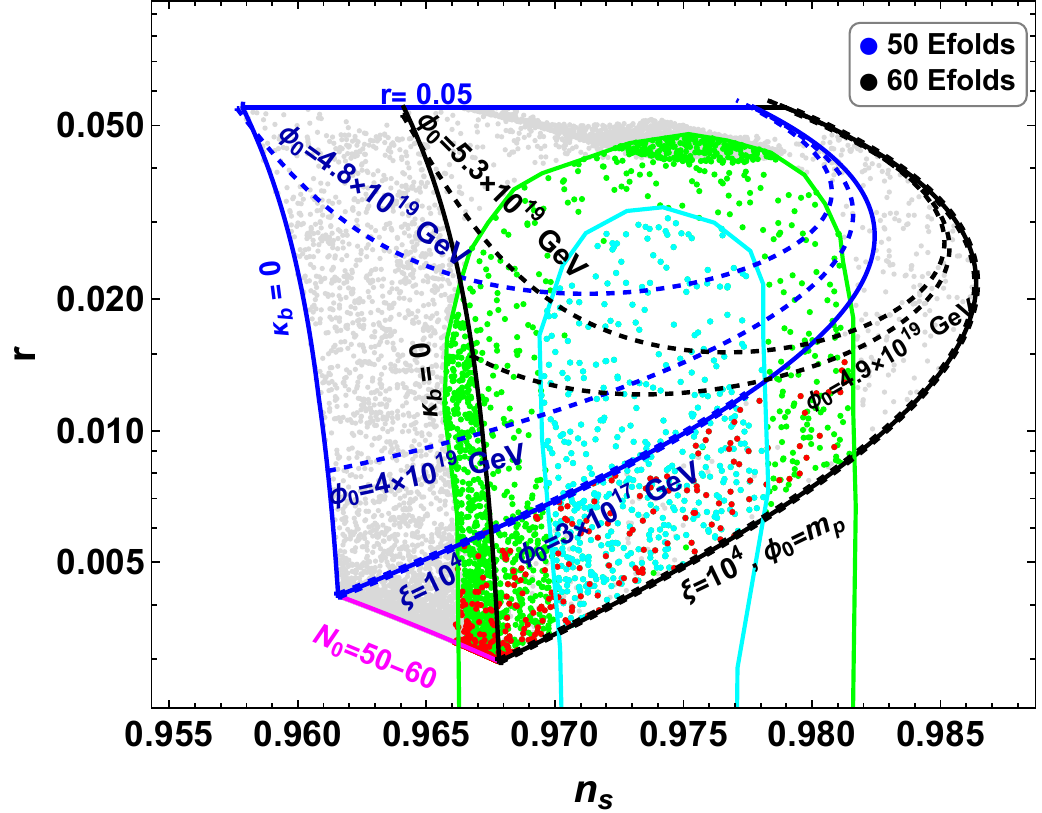}
      \quad
       \caption{The tensor to scalar ratio $r$ is plotted against the scalar spectral index $n_s$, with Cyan and green contours showing the 1\(\sigma\) and 2\(\sigma\) confidence regions from the combined Planck 2018, ACT DR6, and BICEP Keck 2018 (P plus ACT plus LB plus BK18) datasets \cite{ACT:2025tim}. Each panel (a), (b), (c), and (d) shows the inflationary predictions with bosonic corrections for fixed values of $\xi$, $\kappa_b$, $\lambda_{\phi}$, or $\phi_0/m_P$, respectively, while the remaining parameter is varied freely. The plots include predictions for $N_0 = 50$ (blue curves) and $N_0 = 60$ (black curves). Data points are color-coded based on consistency with observational constraints: cyan points fall within the 1\(\sigma\) region, green within 2\(\sigma\), and grey points lie outside the 2\(\sigma\) bounds.}
       \label{fig1}
\end{figure}
\section{Numerical Analytics}\label{sec6}
The inclusion of bosonic and fermionic radiative corrections to the inflationary potential plays a central role in determining the model’s predictions for key cosmological observables, especially the scalar spectral index $n_s$ and the tensor-to-scalar ratio $r$.
Bosonic loop corrections, described by the parameter $\kappa_b$, typically raise the overall height of the potential and make it steeper. This leads to an enhancement in the production of primordial gravitational waves, resulting in a larger value of $r$, while also increasing $n_s$. In this way, bosonic corrections tend to reduce the red tilt of the scalar power spectrum. On the other hand, fermionic corrections, controlled by $\kappa_f$, act in the opposite direction. They flatten the potential, lower the value of $r$, and tend to push $n_s$ toward more red-tilted values.

Recent observational data, particularly from the ACT Data Release 6  \cite{ACT:2025tim} in combination with LB BK18 and Planck 2018, favor a scalar spectral index that is less red-tilted compared to the earlier Planck-only constraints. This trend makes the bosonic corrections more significant for achieving agreement with current data, as they naturally shift the predictions toward higher values of $n_s$ and $r$, offering a better match with the refined measurements.

In the large field regime where mass parameters can be neglected, our model depends on five free parameters: the nonminimal coupling $\xi$, the radiative correction coefficient $\kappa_b $ (or equivalently $\lambda_{\phi H}$), the self-interaction coupling $\lambda_\phi$, and the inflaton field values at the onset and end of inflation, $\phi_0$ and $\phi_e$. These parameters are subject to several theoretical and observational constraints. First, the amplitude of the scalar power spectrum at the pivot scale must match the measured value $A_s(k_0) = 2.19 \times 10^{-9}$, which serves as a key normalization condition for the potential. Second, inflation must end when either of the slow-roll parameters $\epsilon (\phi_e)$ or $\eta (\phi_e)$ reaches unity, setting the value of $\phi_e$. Finally, the field value $\phi_0$ is fixed by requiring a total of $N_0 = 50$ to $60$ e-folds, sufficient to solve the horizon and flatness problems. Together, these constraints restrict the parameter space and ensure the consistency and predictivity of the inflationary framework.

These constraints define the viable parameter space and shape the model’s predictions for inflationary observables. The outcome of our numerical analysis, in the $n_s$–$r$ plane, is presented in Fig.~\ref{fig1} for the inflationary potential including bosonic radiative corrections. The shaded region reflects a random scan over parameters, restricted by the conditions $0.955 < n_s < 0.99$, $50 < N_0 < 60$, and $r < 0.055$. The  cyan  and green contours indicate the 1\(\sigma\) and 2\(\sigma\) confidence levels derived from the combined Planck 2018, ACT DR6, and BICEP Keck 2018 (P plus ACT plus LB plus BK18) datasets \cite{ACT:2025tim}. These contours represent the observational boundaries that any successful inflationary model must satisfy.

In order to better understand the results analytically, we consider an approximate treatment based on Eqs.~(\ref{fullV})--(\ref{delta}). Using these expressions, one can derive various predictions of the radiatively corrected non-minimal \(\phi^4\) inflationary model. Once the parameters \(\xi\), \(\kappa_b\) or \(\kappa_f\), and the number of e-folds \(N_0\) are fixed, the key inflationary observables—namely, the scalar spectral index \(n_s\), and the tensor-to-scalar ratio \(r\), can be computed analytically.
For clarity and completeness, we divide our analysis into three distinct cases:
\begin{enumerate}
    \item \textbf{Case I:} \(\kappa's =0\)  and \(\xi = 0\) — the minimal \(\phi^4\) model.
    \item \textbf{Case II:} \(\kappa's = 0\) and \(\xi \neq 0\) — the tree-level non-minimal \(\phi^4\) model without radiative corrections.
    \item \textbf{Case III:} \(\kappa's \neq 0\) and \(\xi \neq 0\) — the fully radiatively corrected, non-minimal \(\phi^4\) inflation model.
\end{enumerate}
This categorization allows us to isolate the impact of the non-minimal coupling \(\xi\) and the radiative correction parameters \(\kappa_b\), \(\kappa_f\) on the inflationary observables, and to compare their effects in a systematic manner.

In the first case, we consider the tree-level minimal $\phi^4$ inflation model, corresponding to \(\kappa_b = 0, \kappa_f=0\) and \(\xi = 0\), where neither radiative corrections nor non-minimal gravitational coupling is present. The predictions for the inflationary observables in this scenario are well-known and given by:
\begin{align}
n_s & \simeq 1 - 24 \left( \frac{m_P}{\phi_0} \right)^2 \simeq 1 - \frac{3}{N_0}, \quad r \simeq 128 \left( \frac{m_P}{\phi_0} \right)^2 \simeq \frac{16}{N_0}, 
\end{align}
where $\phi_0^2/m_P^2 \simeq 8 N_0 \gg \phi_e^2/m_P^2 \simeq 1/12 $. For a typical choice of \(N_0 = 50-60\), these expressions yield
\[
n_s \simeq 0.94-0.95, \quad r \simeq 0.32-0.26.
\]
These predictions lie well outside the 2\(\sigma\) bounds set by the datasets (Planck + ACT + LB + BK18) \cite{ACT:2025tim}, particularly due to the large tensor-to-scalar ratio \(r\). This discrepancy underscores the need for incorporating radiative corrections and non-minimal couplings, which can significantly alter the inflationary dynamics and bring the theoretical predictions in line with current observational constraints.

To explore Case II, we focus on scenarios with non-minimal coupling (\(\xi \neq 0\)) but no radiative corrections (\(\kappa = 0\)). We analyze this case in two limiting regimes: small \(\xi\) and large \(\xi\).
When the non-minimal coupling is small but nonzero ($\xi \lesssim \mathcal{O}(1/N_0)$), the tree-level predictions of the minimally coupled \(\phi^4\) inflationary model are modified. For \(N_0 = 50\) e-folds, the predictions remain outside the 2\(\sigma\) observational bounds; however, for \(N_0 = 60\) e-folds, the predictions fall within the 2\(\sigma\) region. In this regime, the approximate expressions for the scalar spectral index \(n_s\) and the tensor-to-scalar ratio \(r\) are given by \cite{Okada:2010jf}:
\begin{align}
n_s &\simeq  1 - \frac{3\left(1 + \frac{16}{3} \xi N_0\right)}{N_0\left(1 + 8 \xi N_0\right)}, \quad
r \simeq  \frac{16}{N_0\left(1 + 8 \xi N_0\right)}.
\end{align}
These expressions show that increasing \(\xi\) leads to a suppression of \(r\) and a corresponding enhancement in \(n_s\), thereby shifting the predictions closer to observationally allowed regions. This trend can be seen in Fig.~\ref{fig1} along the line labeled \(\kappa = 0\). Using the recent joint dataset (P+ACT+LB+BK18), the 2\(\sigma\) upper limit for \(n_s\) is \(n_s > 0.9667\). For $N_0=50$ case, the predicted values still lie outside the observational contours. However, for \(N_0 = 60\), the model yields viable solutions within the 2\(\sigma\) confidence interval. From this requirement, we obtain a lower bound on the non-minimal coupling:
\[
\xi \gtrsim  10^{-2}.
\]

In the regime where \(\xi\) is large (\(\xi \gg 1\)), it is convenient to define a dimensionless field variable:
\[
\chi  \equiv \frac{\sqrt{\xi} \phi}{m_P},
\]
such that \(\chi  \gg 1\). In this limit, where $\chi_0^2 \simeq \frac{4N_0}{3} \gg \chi_e^2 \simeq \sqrt{\frac{4}{3}}$, the inflationary predictions for the spectral index and tensor-to-scalar ratio take simple analytic forms:
\begin{align}
n_s &\simeq 1 - \frac{8}{3\chi^2_0} = 1 - \frac{2}{N_0}, \quad
r \simeq \frac{64}{3\chi_0^4} = \frac{12}{N_0^2}.
\end{align}
These predictions, shown as a blue curve labeled $\kappa = 0$ and $N_0 = 60$ in Fig.~\ref{fig1}, indicate that for sufficiently large values of $\xi$, the model yields $n_s$ and $r$ values that fall within the 2\(\sigma\) confidence region of current observational constraints. In contrast, the case with $N_0 = 50$ lies outside the 2\(\sigma\) bounds and is therefore excluded. 
Furthermore, the amplitude of curvature perturbations is approximately given by

\be
A_s \simeq \frac{\lambda_{\phi}}{\xi^2} \left( \frac{\chi^4}{768 \pi^2} \right) 
\simeq \frac{\lambda_{\phi}}{\xi^2} \left( \frac{N_0^2}{432 \pi^2} \right),
\ee
which leads to the relation,
\be 
\xi \simeq \sqrt{\lambda_{\phi}} \left( \frac{N_0}{50} \right) \times 10^4.
\ee
To realize inflation driven by the Standard Model Higgs boson, one typically requires a quartic coupling $\lambda_H \sim 1$, which in turn demands a large nonminimal coupling $\xi_H \sim 10^4$. While this setup is consistent with the measured Higgs mass of 125~GeV, it demands that $\lambda_H$ remain positive at high field values and generally predicts a scalar spectral index $n_s$ that is in tension with recent observations from the Atacama Cosmology Telescope (ACT). As we discuss below, our model avoids these shortcomings.

In the third case, where both the non-minimal coupling \(\xi\) and the radiative corrections (parametrized by \(\kappa_b\)) are nonzero, the inflationary predictions receive significant modifications compared to the tree-level results. Here, we focus on the bosonic contributions to the effective potential. The inclusion of one-loop corrections alters the dynamics of the inflaton field and leads to the following approximate expressions for the inflationary observables \cite{Okada:2010jf}:
\bea
n_s &\simeq& 1 - \frac{8}{3\chi^2} \left( \frac{1 - \frac{6\kappa_{b}}{\lambda_{\phi}}(3 + 4\ln(\sqrt{\xi}\chi))}
{1 - \frac{24\kappa_{b}}{\lambda_{\phi}} \ln(\sqrt{\xi}\chi)}\right), \\
r &\simeq&   \frac{64}{3\chi^4} \left( \frac{1 + \frac{6\kappa_{b}}{\lambda_{\phi}}(\chi^2 - 4\ln(\sqrt{\xi}\chi))}
{1 - \frac{24\kappa_{b}}{\lambda_{\phi}} \ln(\sqrt{\xi}\chi)}\right)^2.
\eea
The amplitude of the curvature perturbations is given by:
\begin{equation}
A_s \simeq 
\frac{\lambda_{\phi}}{\xi^2} \left( \frac{\chi^4}{768\,\pi^2} \right)
\frac{\left( 1 - \frac{24\kappa_{b}}{\lambda_{\phi}} \ln(\sqrt{\xi}\chi) \right)^3}
{\left(1 - \frac{6\kappa_{b}}{\lambda_{\phi}}(\psi^2 + 4\ln(\sqrt{\xi}\chi))\right)^2}.   
\end{equation}
The impact of these radiative corrections is clearly illustrated in Figs.~\ref{fig1}, as shown by the dashed lines. Notably, the inclusion of \(\kappa_{b}\) results in an enhancement of both the spectral index \(n_s\) and the tensor-to-scalar ratio \(r\), compared to the tree-level non-minimal case. This increase can be attributed to the loop-induced modifications of the inflaton potential, which effectively alter the inflationary dynamics and shift the predictions towards regions favored by current observational bounds. 

\begin{figure}[t]
\centering
\includegraphics[width=0.48\linewidth,height=6.7cm]{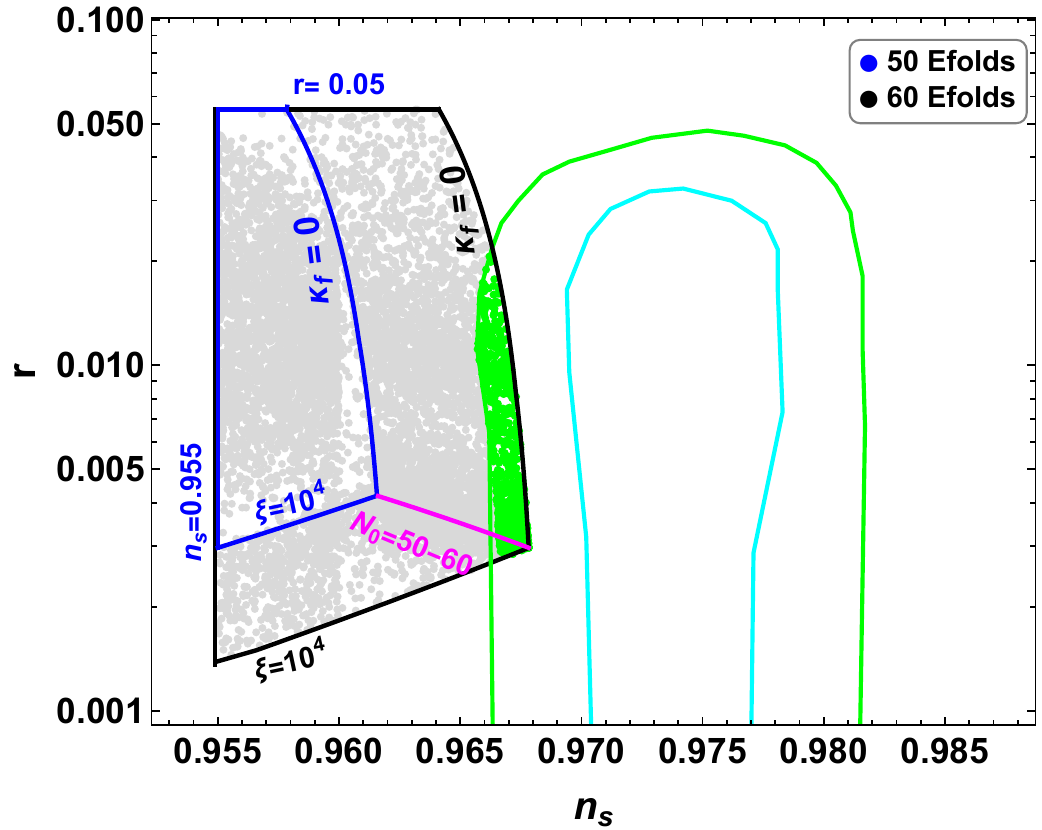}
\quad
\includegraphics[width=0.48\linewidth,height=6.7cm]{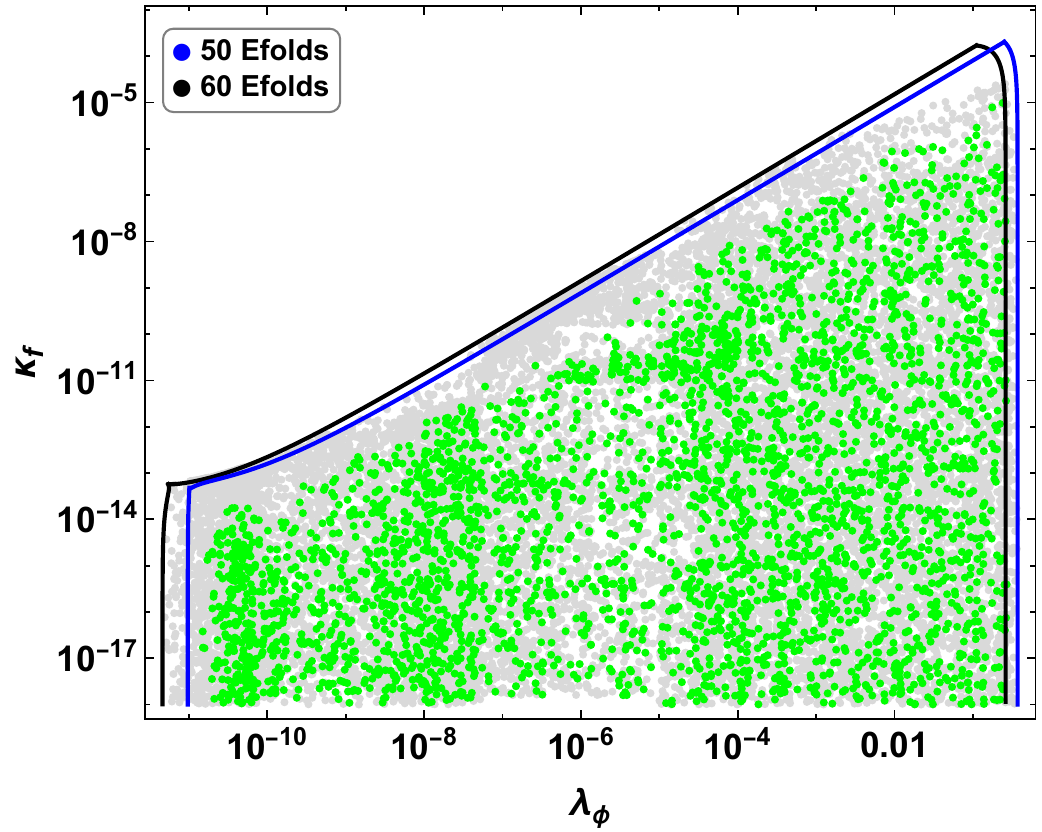}
\quad
\caption{The tensor to scalar ratio $r$ is plotted against the scalar spectral index $n_s$, with cyan and green contours showing the 1\(\sigma\) and 2\(\sigma\) confidence regions from the combined Planck 2018, ACT DR6, and BICEP Keck 2018 (P plus ACT plus LB plus BK18) datasets \cite{ACT:2025tim}. The inflationary predictions with fermionic corrections are shown for $N_0 = 50$ (blue curves) and $N_0 = 60$ (black curves). Data points are color-coded based on consistency with observational constraints: green points fall within the 2\(\sigma\) and grey points lie outside the 2\(\sigma\) bounds.}
\label{fig2}
\end{figure}

We have demonstrated that bosonic corrections align well with the observational data from ACT. For completeness, we also consider the scenario in which fermionic corrections dominate over bosonic ones. As illustrated in Fig.~\ref{fig2}, this setup accounts for a limited portion of the $2\sigma$ confidence region in the $r$–$n_s$ plane, as constrained by the combined ACT and related datasets. However, it is important to note that fermionic contributions can still play a crucial role, especially in the context of reheating and baryogenesis via leptogenesis. For a detailed discussion on the fermionic case, we refer the reader to Okada et al.~\cite{}, where the implications for post-inflationary dynamics and leptogenesis are explored in depth.

\section{RGE-Improved Treatment for Dark Matter Inflaton Scenario}\label{sec7}
To illustrate a realistic example, we consider a scenario in which the inflaton $\phi$ also serves as a dark matter candidate \cite{Lerner:2009xg, Okada:2010jd}, with a mass around $m \simeq 1$~TeV. Imposing a $Z_2$ symmetry under which $\phi \rightarrow -\phi$ eliminates the Yukawa interaction involving the $y_{\phi}$ coupling, thereby stabilizing $\phi$ and making it a viable dark matter particle. In this framework, the inflaton couples to the Standard Model Higgs via the portal interaction $\lambda_{\phi H} \phi^2 H^\dagger H$, which is tightly constrained by dark matter relic density and direct detection limits to be approximately $\lambda_{\phi H} \sim 0.1$ \cite{Cirelli:2024ssz,Bharadwaj:2024crt}. We analyze this scenario using a renormalization group improved approach, as detailed below.

In the Einstein frame, the RGE-improved scalar potential takes the following form,
\begin{equation}
V_E(\phi) = \frac{\frac{1}{4!} \lambda_{\phi}(t)\,G(t)^4\,\phi^4}{\left(1+\frac{\xi(t)\,G(t)^2\phi^2}{m_P^2}\right)^2},
\label{potrgi}
\end{equation}
where $t=\ln(\phi/\mu)$, and $G(t) = \exp\left(- \int_0^t \frac{\gamma(t')}{1+\gamma(t')} dt'\right) \approx 1$, with $\gamma(t) \approx 0$ denoting the anomalous dimension of the gauge singlet inflaton. For the scalar dark matter scenario, the relevant RGEs are \cite{Lerner:2009xg}:
\bea
\frac{d\lambda_{\phi}}{dt} &=& \frac{1}{(4\pi)^2} \left( 3 s_{\phi}^2 \lambda_{\phi}^2  + 48 \lambda_{\phi H}^2 \right), \label{eq:rge_lambda} \\[10pt]
\frac{d\lambda_{\phi H}}{dt} &=& \frac{ \lambda_{\phi H}}{(4\pi)^2} \left( 8 s_{\phi} \lambda_{\phi H} + \lambda_{\phi} + 6 \lambda_{H}  - \frac{9}{10}(g_1^2 +  5 g_2^2) + 6 y_t^2 \right), \label{eq:rge_g}
\\[10pt]
\frac{d \xi}{dt} & \simeq & \frac{ \xi}{(4\pi)^2} \left( 6 s_{\phi} \lambda_{\phi H}  \right), \quad \frac{d \xi_H}{dt}  \simeq  \frac{ \xi}{(4\pi)^2} \left( s_{\phi} \lambda_{\phi H}  \right),   \label{eq:rge_xiphi}
\eea
assuming $\xi \gg 1 \gtrsim \xi_H $. The SM RGE for SM Higgs quartic coupling $\lambda_H$ receive the following contribution from $\lambda_{\phi}$ and $\lambda_{\phi H}$ at 1-loop level:
\bea
\frac{d \lambda_H}{dt} &=& \frac{1}{(4\pi)^2} \Bigg( 
24 \lambda_H^2 - 6 y_t^4 + \left( \frac{3}{8} g_1^4 + \frac{3}{4} g_1^2 g_2^2 + \frac{9}{8} g_2^4 \right) \nonumber \\
& + & \left( -9 g_2^2 - 3 g_1^2 + 12 y_t^2 \right) \lambda_H + 2 s_\phi^2 \lambda_{\phi H}^2  \Bigg),
\eea
where the suppression factor $s_\phi$ is given by 
\be
s_\phi (t) = \frac{1 + \left( \dfrac{\xi \mu^2 e^{2t}}{m_P^2} \right)}{1 + (6\xi + 1)\left( \dfrac{\xi \mu^2 e^{2t}}{m_P^2} \right)}. 
\ee
In contrast, the corresponding suppression factor for the Standard Model Higgs field, $s_H$, effectively reduces to unity since the Higgs remains stabilized at the origin during inflation. Moreover, in the regime where $s_\phi \ll 1$ and $\lambda_{\phi H} \lesssim 1$, the nonminimal couplings exhibit negligible running and can be treated as approximately scale-independent.

In our numerical analysis, we solve the full set of two-loop RGEs for SM gauge couplings $g_1, g_2, g_3$, the top Yukawa coupling $y_t$, and the Higgs quartic coupling $\lambda_H$, incorporating the one-loop level modifications due to the extended scalar sector. These are supplemented with the one-loop RGEs for the portal coupling $\lambda_{\phi H}$ and the inflaton self-coupling $\lambda_{\phi}$. For the numerical implementation, we fix the renormalization scale $\mu$ at the top quark pole mass $M_t$. To maintain the validity of the effective field theory, we require the inflationary scale to remain below the cutoff scale $\Lambda = m_P$ \cite{Bezrukov:2010jz}.

\begin{table}[t]
\centering
\begin{tabular}{|c|c|c|c|c|c|c|c|c|c|c|c|c|}
\hline
$\phi_0\,(\mathrm{GeV})$ & $\phi_e\,(\mathrm{GeV})$ & $N$ & $r$ & $n_s$ & $-\alpha_s$ & $\xi$ & $\lambda_\phi(0)$ & $\lambda_{\phi H}(0)$ & $\lambda_{\phi}(t_0)$&$\lambda_{\phi H}(t_0)$ & $\lambda_{H}(t_0)$ & $s_{\phi}(t_0)$ \\
\hline
$2.3 \times 10^{17}$ & $2.6 \times 10^{16}$ & $50.$ & $0.006$ & $0.970$ & $7\times 10^{-4}$ & $10^4$ & $0.14$ & $0.14$& $0.55$&$0.22$ & $0.049$ & $1.7 \times 10^{-5}$  \\
\hline
$2.3 \times 10^{17}$ & $2.6 \times 10^{16}$ & $60$ & $0.005$ & $0.975$ & $5 \times 10^{-4}$ & $10^4$ & $0.12$ & $0.12$& $0.40$ & $0.18$ & $0.062$ & $1.7 \times 10^{-5}$ \\
\hline
$3.1 \times 10^{17}$ & $3.7 \times 10^{16}$ & $50$ & $0.005$ & $0.97$ & $7 \times 10^{-4}$ & $5 \times 10^3$ & $0.06$ & $0.06$& $0.11$ & $0.07$ & $0.016$ & $3.4 \times 10^{-5}$ 
\\
\hline
$3.4 \times 10^{17}$ & $3.7 \times 10^{16}$ & $60$ & $0.004$ & $0.972$ & $5 \times 10^{-4}$ & $5 \times 10^3$ & $0.05$ & $0.05$& $0.08$ & $0.06$ & $0.011$ & $3.4 \times 10^{-5}$ 
\\
\hline
\end{tabular}
\caption{Benchmark values of the inflationary observables computed using the renormalization group improved effective potential given in Eq.~(\ref{potrgi}). The variable $t_0$ is defined as $t_0 \equiv \ln(\phi_0/\mu)$.}
\label{tab:inflation-benchmark-full}
\end{table}

The benchmark points shown in Table~\ref{tab:inflation-benchmark-full} represent a viable solution within the framework of the renormalization group–improved inflationary scenario. The resulting tensor-to-scalar ratio, $r \approx 0.004{-}0.006$, and scalar spectral index, $n_s \approx 0.970-0.975$, for the number of e-folds $N_0 \approx 50{-}60$, are consistent with the latest observational constraints from ACT, Planck, and BICEP/Keck. Notably, the predicted range for $r$ lies within the sensitivity reach of upcoming CMB polarization experiments \cite{LiteBIRD:2022cnt, CMB-S4:2020lpa,SimonsObservatory:2018koc}. All inflationary observables are computed under the assumption of sub-Planckian inflaton field values, $\phi < m_P$, during inflation. The Higgs portal coupling $\lambda_{\phi H}$ contributes positively to the running of the Standard Model Higgs quartic coupling $\lambda_H$, helping to stabilize the Higgs field at the origin during inflation. As shown in Table\~I, this results in a positive value $\lambda_H \simeq 0.01-0.06$ at the relevant inflationary scale. Furthermore, the allowed range of the Higgs portal coupling $\lambda_{\phi H}$ remains compatible with the observed dark matter relic abundance and current direct detection constraints \cite{Cirelli:2024ssz, Bharadwaj:2024crt}

Assuming that $d\lambda_{\phi}/dt$ remains approximately constant and that $s_{\phi} \lambda_{\phi} \ll \lambda_{\phi H}$, the effective potential in the Einstein frame can be approximated as
\begin{equation}
V_E(\phi) \simeq \frac{\frac{1}{4!}\left(\lambda_0 
+ \frac{48}{(4\pi)^2}   \lambda_{\phi H}^2 \ln \left( \phi / \mu \right) \right)  \phi^4}
{\left(1+\frac{\xi\,\phi^2}{m_P^2}\right)^2}, 
 \label{ApproxPotential}
\end{equation}
where $\lambda_0 = \lambda_{\phi}(t = 0)$ denotes the initial value of the inflaton quartic coupling. This expression provides a reliable approximation in the relevant parameter space and identifies the effective bosonic correction coefficient $\kappa_b = 2 \lambda_{\phi H}^2 / (4\pi)^2$. These results are in good agreement with the earlier numerical analysis carried out for a more general setup.
\section{Conclusion}
\label{sec:conclusion}

We explored an inflationary framework based on a quartic scalar potential along with radiative corrections and a nonminimal coupling to gravity. We started from the Jordan frame and constructed a general scalar field theory that includes interactions with both bosonic and fermionic sectors. A conformal transformation to the Einstein frame helped us obtain a canonically normalized inflaton field and calculate the quantum-corrected effective potential, incorporating the main one-loop contributions.

We showed that the minimally coupled quartic inflation model does not fit current cosmological observations, particularly those from the Planck 2018 and ACT DR6 datasets. This is due to its predictions for the scalar spectral index and tensor-to-scalar ratio. However, adding a small nonminimal coupling of $\xi \gtrsim 0.02$ with $N_0 \sim 60$ is enough to align the model with observations, creating a narrow but viable parameter space window.
Our results highlight the significance of bosonic radiative corrections, which tend to raise the scalar spectral index and provide predictions that align more closely with recent ACT data. In contrast, fermionic corrections usually create a more red-tilted spectrum and suppress tensor modes, making them less favorable in today's observational landscape.

We also looked into scenarios where the inflaton field could act as a dark matter candidate, especially through a Higgs portal interaction. This offers an intriguing chance to combine inflation and dark matter within a single scalar sector, with predictions that can be tested in future cosmological and direct detection experiments.

\bibliographystyle{apsrev4-1}

\bibliographystyle{unsrt}  
\bibliography{References}  

\end{document}